\documentstyle[12pt]{article}
\newdimen\underbracklength
\newcommand\underbrack[1]{
  \setbox0=\hbox{$#1$}\underbracklength=\wd0\advance\underbracklength by-5pt
  \setbox0=\hbox{\kern2pt\vrule width.5pt height12pt depth-8pt
    \vbox{\hrule width\underbracklength height8.5pt depth-8pt}%
    \vrule width.5pt height12pt depth-8pt\kern2pt}
  \vtop{\hbox{$#1$}\box0\kern-4pt}}
\catcode`@=11
\newcount\@tempcntc
\def\@citex[#1]#2{\if@filesw\immediate\write\@auxout{\string\citation{#2}}\fi
  \@tempcnta\z@\@tempcntb\m@ne\def\@citea{}\@cite{\@for\@citeb:=#2\do
    {\@ifundefined
       {b@\@citeb}{\@citeo\@tempcntb\m@ne\@citea\def\@citea{,}{\bf ?}\@warning
       {Citation `\@citeb' on page \thepage \space undefined}}%
    {\setbox\z@\hbox{\global\@tempcntc0\csname b@\@citeb\endcsname\relax}%
     \ifnum\@tempcntc=\z@ \@citeo\@tempcntb\m@ne
       \@citea\def\@citea{,}\hbox{\csname b@\@citeb\endcsname}%
     \else
      \advance\@tempcntb\@ne
      \ifnum\@tempcntb=\@tempcntc
      \else\advance\@tempcntb\m@ne\@citeo
      \@tempcnta\@tempcntc\@tempcntb\@tempcntc\fi\fi}}\@citeo}{#1}}
\def\@citeo{\ifnum\@tempcnta>\@tempcntb\else\@citea\def\@citea{,}%
  \ifnum\@tempcnta=\@tempcntb\the\@tempcnta\else
   {\advance\@tempcnta\@ne\ifnum\@tempcnta=\@tempcntb \else \def\@citea{--}\fi
    \advance\@tempcnta\m@ne\the\@tempcnta\@citea\the\@tempcntb}\fi\fi}
\catcode`@=12
\begin{document}

\thispagestyle{empty}
\begin{flushright}
MZ-TH/95-33 \\[-0.2cm]
December 1995 \\[-0.2cm]
\end{flushright}

\begin{center}
{\Large\bf\boldmath{$\Lambda_b$}-Polarization from its Inclusive
  Semileptonic Decay}%
  \footnote{Invited talk given at the conference ``Production and Decay of
  Hyperons, Charm and Beauty\\\hbox{\qquad} Hadrons'', Strassbourg, France,
  September 5--8, 1995; to appear in the Proceedings\\}
\vspace{1truecm}\\
{\large J\"urgen G.~K\"orner%
  \footnote{Supported in part by the BMBF, FRG, under contract 06MZ566,\\
  \hbox{\qquad}and by HUCAM, EU, under contract CHRX-CT94-0579}}
\vspace{0.4truecm}\\
Institut f\"ur Physik, Johannes Gutenberg-Universit\"at\\
Staudingerweg 7, D-55099 Mainz, Germany
\end{center}
\vspace{1truecm}
\centerline{\bf ABSTRACT}\noindent
We discuss the $\Lambda_b$-polarization of the bottom baryon $\Lambda_b$ in
its production and in its decay. Standard Model predictions for $b$-quark
polarization on the $Z$-peak are reviewed, including beam-quark orientation
effects and $O(\alpha_s)$ radiative corrections to the polarization. The
$b$-quark polarization is expected to be handed over to the $\Lambda_b$
into which it fragments with a high degree of efficiency. We present several
spin-momentum correlation measures of the spin of the $\Lambda_b$
with the momenta of its decay products in its inclusive semileptonic decay.
We estimate the strengths of the spin-momentum correlations in the
Free Quark Decay (FQD) model and find that some of the measures provide
good or very good analyzing power for measuring the polarization of
the $\Lambda_b$.

\newpage

\section{Introduction}
This presentation will cover four different topics, all relevant for the
polarization of $\Lambda_b$'s from Z-decays. Interest in the subject
of $\Lambda_b$-polarization was sparked by a recent LEP measurement of
$\Lambda_b$-polarization by the ALEPH-collaboration, which found a rather
small value for the $\Lambda_b$-polarization, smaller than SM-expectations
by about one standard deviation~\cite{strass1}.

The first topic concerns the polarization of $b$-quarks produced on the
$Z$-peak which turns out to be rather large ($93\%$) in the Standard
Model (SM). Recently the $O(\alpha_s$) corrections to the polarization of
the $b$-quarks have been calculated~\cite{strass2,strass3,strass4}. The
O($\alpha_s$) corrections to the Born term values for the polarization
turn out to be rather small. The second topic concerns the polarization
transfer of the polarization of the $b$-quarks to the $\Lambda_b$'s which
can be expected to be rather efficient in as much as the polarization
transfer is 100\% in the heavy quark limit~\cite{strass5}. In the third
part we briefly discuss the description of inclusive semileptonic
$\Lambda_b$ decays using the operator product expansion (OPE) method
within Heavy Quark Effective Theory (HQET)~\cite{strass6,strass7}. The main
result is that the leading contribution to the decay is given by the
usual free quark decay (FQD) amplitude with small $O(1/m_b^2$) corrections.
In the last part we give a brief discussion of how the $\Lambda_b$
polarization was determined by the ALEPH experiment and suggest
improvements on the ALEPH measurement. In particular we propose to measure
the $\Lambda_b$ polarization through spin-momentum correlations of the
spin of the $\Lambda_b$ with the momenta of its decay products.

\section{$b$-Quark Polarization on the $Z$-Peak}
Because of the small value of $\sin^2\theta_W$ the neutral current coupling
of the $Z$ to quark pairs is dominantly left-chiral. Also, since the quark
masses are small on the scale of the $Z$-mass, quark mass effects can
practically be neglected in the production process, and thus the produced
quarks are predicted to  emerge strongly polarized. The polarization effect
is strongest for the $T_3=-1/2$ ($Q=-1/3$) quarks with 93\% left-handed
polarisation, and somewhat reduced for the $T_3=1/2$ ($Q=2/3$) with 60\%
left polarization. The polarization is mainly longitudinal with small
transverse polarization components. The latter are suppressed by the factor
$m_b/m_Z$ and thus small.

In Fig.~1 we plot the $\cos\theta$-dependence of the alignment
polarization $P^\ell$ (often also called the longitudinal polarization) of
the $b$-quark on the $Z$-peak. The angle $\theta$ is the polar angle of the
quark relative to the $e^-$-beam direction. The alignment polarisation
$P^\ell(cos\theta)$ shown in Fig.~1 is large and negative and shows
very little $\cos\theta$-dependence. Note that the $O(\alpha_s)$ radiative
corrections are small. They never amonunt to more than 2\% over the whole
$\cos\theta$-range. As mentioned before the transverse components are quite
small and can be neglected for all practical purposes, in particular when
they are averaged over the relative beam-quark orientation~\cite{strass3}.

\section{Polarization Transfer $b^{(\uparrow)}\rightarrow
  \Lambda_b^{(\uparrow)}$}
Given that the $b$-quark from $Z$-decays has a large negative longitudinal
polarization one has to face the issue of how efficiently this polarization
is transferred to the $\Lambda_b$. Suppose first that the polarized $b$-quark
picks up a spin-0 $ud$ pair to form a `prompt' $\Lambda_b$. Due to its
$ud$ pair having spin 0, all of the $\Lambda_b$ spin resides on the valence
$b$-quark and we expect $b$ polarization to become $\Lambda_b$-polarization
(in the heavy quark limit where $b$ spin-flip is suppressed during
hadronization). Suppose instead that the polarized $b$-quark had combined
with a spin-1 $ud$ pair to form a $\Sigma_b$ or a $\Sigma_b^*$. In the heavy
quark limit these two states would form a degenerate pair of states, i.e.
the $b$-quark polarization would again be handed down to the $\Lambda_b$
when the coherent superposition of the two $\Sigma_b$ and $\Sigma_b^*$ states
decay into the $\Lambda_b$. The same reasoning applies to any of the other
degenerate excited Heavy Quark Symmetry doublets decaying into the
$\Lambda_b$. The conclusion is that the $b$-quark polarization is
transferred to the $\Lambda_b$  with 100\% efficiency in the heavy quark
limit~\cite{strass5}. In real life the Heavy Quark Symmetry doublets are
not quite degenerate which can be appreciated to lead to a reduction in the
efficiency factor of the polarization transfer by the above line of
reasoning~\cite{strass5,strass8}. Falk and Peskin have attempted to
estimate the reduction factor and conclude that one retains an efficiency
factor of $\cong 75$\% in the polarization transfer from $b$-quarks to
$\Lambda_b$'s~\cite{strass8}.

\section{Theoretical Framework for\\ Inclusive Semileptonic Decays}
Next we turn to the description of the semileptonic inclusive decays of the
$\Lambda_b$. Recently there have been important advances in our
understanding of these decays using OPE and HQET
techniques~\cite{strass6,strass7}. The heavy mass scale in these decays
allows one to perform an OPE-expansion of the bilocal product of weak
currents appearing in the rate expressions. This procedure is quite similar
in spirit to the OPE-expansion in deep inelastic scattering where $q^2$
sets the large energy scale. Matrix elements of the operators appearing in
the OPE-expansion are parametrized in terms of a set of basic
nonperturbative parameters using HQET-methods. An important result of this
analysis is that to lowest order in the heavy quark mass expansion one
recovers the usual parton model result corresponding to Free Quark Decay
(FQD). Corrections to FQD start appearing only at $O(1/m_b^2)$ and are
therefore small. Because of the fact that the O$(1/m_b^2)$ corrections are
so small we shall, for the sake of simplicity, base our following discussion
on the FQD results. A treatment of the $O(1/m_b^2)$ and
$O(\alpha_s)$~\cite{strass9} corrections can be found in~\cite{strass10}.

The first experimental determination of the $\Lambda_b$ polarization
by the ALEPH-collaboration gave the intriguingly small value
\hbox{$P\!=\!-0.23^{+0.24}_{-0.20}\pm 0.08$} \cite{strass1}. The measurement
was based on a proposal by Bonvicini and Randall to determine the $\Lambda_b$
polarization by a measurement of the ratio of the average of the electron
and neutrino energies $y=\langle E_e\rangle/\langle E_\nu\rangle$~%
\cite{strass14}. Compared to the well-tested method of determining the
muon's polarization from the shape of the electron spectrum in its
semileptonic decay one is here suffering from the fact that the momentum of
the $\Lambda_b$ is practically not known in the corresponding semileptonic
$\Lambda_b$ decay, i.e. there is no unpolarized reference spectrum to
compare with. The advantage of the Bonvicini-Randall measure is that any
fragmentation dependence practically drops out of the ratio of averages of
electron and neutrino energies.

The polarization of the $\Lambda_b$ directly affects the spectra of the
elctron (or muon) and the $\nu$ produced in its semileptonic decays.
The qualitative nature of the change can be obtained by an examination of
the angular decay distributions of the leptons in the rest frame of the
decaying $\Lambda_b$. The electrons (or muons) tend to be emitted
antiparallel to the spin of the $\Lambda_b$ and the $\nu$'s are emitted
preferentially parallel to its spin~\cite{strass9}. In the lab frame this
translates into a harder electron spectrum and a softer neutrino spectrum
as compared to the unpolarized case. It goes without saying that this is a
difficult measurement since one needs to reconstruct the neutrino energy in
the decays. In fact the difficulty in the reconstruction of the neutrino
energy is the main source of the rather large errors quoted in the
measurement of the ALEPH-collaboration. Possible optimizations of
spectra-related measures have been investigated in~\cite{strass10} where
the measure $y_2=\langle E_e^2\rangle/\langle E_\nu^2\rangle$ was found to
be a good candidate to reduce the errors in this measurement.

\section{Spin-Momentum Correlations in Inclusive Semileptonic Decays of
Polarized~$\Lambda_b$'s}
In Ref.\cite{strass11} we investigated the potential of using spin-momentum
correlations between the spin of the $\Lambda_b$ and the momenta of its
decay products to determine the polarization of the $\Lambda_b$. Consider
a polarized $\Lambda_b$ in its rest frame situated at the origin of a
coordinate system with its polarization vector pointing into the
$(\theta,\phi)$-direction as depicted in Fig.~2 and Fig.~3.
In Fig.~2 the $z$-axis is in the $(x,z)$-decay plane. Such a
coordinate system is referred to as a helicity system. In Fig.~3
the $z$-axis is perpendicular to the decay plane, which now is in the
$(x,y)$-plane. Such a system will be referred to as a transversity system.
The orientation of the polarization vector is given in terms of the polar
angle $\theta$ and $\phi$ in the helicity system and in terms of $\theta_T$
and $\phi_T$ in the transversity system.

We begin our discussion with the helicity system. The two-fold angular
decay distribution in the helicity system is given by the
expression~\cite{strass11,strass12,strass13}
\begin{equation}\label{eqn1}
\frac{d\Gamma}{d\cos\theta\,d\phi}=A+P(B\cos\theta+C\sin\theta\cos\phi),
\end{equation}
where the polarization is now defined as $P=|\vec P|$. The values of the
angular coefficients $A$, $B$ and $C$ depend on the decay dynamics and
on whether the momentum of the electron (system I) or the momentum of the
recoiling hadron system (system II) is used to define the $z$-direction in
the helicity system. Instead of analyzing the two-fold decay distribution
we reduce the decay distribution Eq.~(\ref{eqn1}) to single angle decay
distributions by the appropiate $\theta$ and $\phi$ integrations. We obtain
\begin{eqnarray}
\frac{d\Gamma}{d\cos\theta}&\propto&1+\alpha_PP\cos\theta\\
\frac{d\Gamma}{d\phi}&\propto&1+\gamma_PP\cos\phi
\end{eqnarray}
where, in terms of the angular coefficients $A$, $B$ and $C$ defined above,
the asymmetry parameters are given by $\alpha_P=B/A$ and $\gamma_P=\pi C/4A$.

In Table~\ref{tab1} we list the values of the asymmetry parameters in the
FQD model for $b\rightarrow c$ and $b\rightarrow u$ transitions in system I
and II, respectively. The asymmetry measures are large enough to provide
useful measures for the determination of the polarization of the $\Lambda_b$.
The azimuthal asymmetry tends to be somewhat larger than the polar asymmetry.
For the case $b\rightarrow c$ the polar asymmetry $\alpha_P$ in system I
(where the electron defines the $z$-axis) is smaller than that in system II
(see also~\cite{strass9}). The azimuthal asymmetry is approximately the same
in both systems and is quite close to its mass zero value $\gamma_P=-0.56$.
For the decay into a mass zero quark as the $u$-quark the asymmetry values
in system I and II coincide. This is a simple consequence of the fact that
the $u$-quark and the electron are Fierz symmetry partners in the decay. In
the case of mass degeneracy as is the case here the decay distributions are
symmetric under the exchange of the two.

In order to draw the analogy to muon decay we arrange the decay products
in muon decay in the same weak isospin order as in the $b\rightarrow u$
case. One has
\begin{equation}
b\rightarrow\underbrack{u+e^-}+\bar\nu_e,\qquad
\mu^-\rightarrow\underbrack{\nu_\mu+e^-}+\bar\nu_e,
\end{equation}
where we have added the braces connecting the Fierz partners for emphasis.
The result for the polar asymmetry parameter $\alpha_P=-1/3$ is well
familiar from muon decay when the mass of the electron is neglected. The
corresponding value $\gamma_P=-2\pi^2/35$ for the azimuthal asymmetry
parameter has not been widely publicized in the muon decay case for the
obvious reason that the azimuthal asymmetry cannot be measured with two
undetected neutrinos in the final state.

Next we turn to the transversity system as drawn in Fig.~3. The
relation between the transversity angles ($\theta_T$, $\phi_T$) and the
helicity angles ($\theta$, $\phi$) can be easily seen to be given by
\begin{eqnarray}
\cos\theta_T&=&\sin\theta\sin\phi\nonumber\\
\sin\theta_T\sin\phi_T&=&\sin\theta\cos\phi\\
\sin\theta_T\cos\phi_T&=&\cos\theta.\nonumber
\end{eqnarray}
Correspondingly one has the two-fold angular decay distribution
\begin{equation}\label{eqn2}
\frac{d\Gamma}{d\cos\theta_Td\phi_T}
  =A+P\sin\theta_T(B\cos\phi_T+C\sin\phi_T)
\end{equation}
Integrating Eq.~(\ref{eqn2}) over $\phi_T$ leads to a flat
$\cos\theta_T$-distribution while integrating over $\cos\theta_T$ one
obtains the one-fold angular decay distribution
\begin{equation}
\frac{d\Gamma}{d\phi_T}\propto 1+\gamma_P^TP\cos(\phi_T-\beta),\quad
  \mbox{where}
\end{equation}
\begin{equation}
\gamma_P^T=\frac{\sqrt{B^2+C^2}}{4A}\,\pi
\end{equation}
and the phase angle $\beta$ is given by
\begin{equation}
\beta=\arcsin\,\frac{C}{\sqrt{B^2+C^2}}.
\end{equation}
In Table~\ref{tab2} we list the FQD values of the asymmetry parameter
$\gamma_P^T$ and the phase angle $\beta$, again for $b\rightarrow c$ and
$b\rightarrow u$ transitions in the two respective systems I and II. The
asymmetry parameter is large and negative in both systems.

As in the case of the helicity systems the values of the asymmetry
parameter $\gamma_P^T$ and the phase angle $\beta$ agree for systems I and
II for the $b\rightarrow u$ transitions. In fact, mass effects of the final
quark play no important role altogether as can be seen by comparing the
$b\rightarrow c$ case with the $b\rightarrow u$ case.

In conclusion the polarization measures discussed in this talk should prove
useful in analyzing the polarization of $\Lambda_b$-quarks. They complement
the spectrum related measures discussed in~\cite{strass10,strass14}.
\vspace{0.2cm}\\
\noindent{\bf Acknowledgement:} I would like to thank my collaborators
C.~Diaconu, S.~Groote, D.~Pirjol, M.~Talby and M.M.~Tung for their
participation in this work and F.~Scheck for a discussion on polarization
effects in $\mu$-decay.

\vspace{1truecm}

\centerline{\bf\Large Table Captions}
\vspace{0.5truecm}
\newcounter{tabcap}
\begin{list}{\bf\rm Tab.\ \arabic{tabcap}: }{\usecounter{tabcap}
\labelwidth1.6cm \leftmargin2.5cm \labelsep0.4cm \itemsep0ex plus0.2ex }

\item Values of polar and azimuthal asymmetry in the Free Quark Decay
model in system I and II (with $m_b=4.81$ GeV, $m_c=1.45$ GeV, $m_u=0$).
Also shown is numerical value of the factor $-2\pi^2/35$.

\item Values of azimuthal asymmetry parameter $\gamma_P^T$ and phase
angle $\beta$ in the FQD model in system I and II

\end{list}

\vspace{1truecm}

\hbox{\vbox{\hsize=9cm
\centerline{\begin{tabular}{|l||l|l|}
\hline
&$\alpha_P$&$\gamma_P$\\
\hline\hline
System I&&\\
$b\rightarrow c$&$-0.26$&$-0.59$\\
$b\rightarrow u$&$-1/3$&$-2\pi^2/35\ (=-0.56)$\\
\hline
System II&&\\
$b\rightarrow c$&$-0.42$&$-0.54$\\
$b\rightarrow u$&$-1/3$&$-2\pi^2/35\ (=-0.56)$\\
\hline
\end{tabular}}
\vspace{0.5truecm}
\centerline{\bf\Large Table 1\label{tab1}}}
\hfill
\vbox{\hsize=7cm
\centerline{\begin{tabular}{|l||l|l|}
\hline
&$\gamma_P^T$&$\beta$\\
\hline\hline
System I&&\\
$b\rightarrow c$&$-0.62$&$71^0$\\
$b\rightarrow u$&$-0.62$&$65^0$\\
\hline
System II&&\\
$b\rightarrow c$&$-0.63$&$59^0$\\
$b\rightarrow u$&$-0.62$&$65^0$\\
\hline
\end{tabular}}
\vspace{0.5truecm}
\centerline{\bf\Large Table 2\label{tab2}}}}

\vspace{1truecm}

\centerline{\bf\Large Figure Captions}
\vspace{0.5truecm}
\newcounter{figcap}
\begin{list}{\bf\rm Fig.\ \arabic{figcap}: }{\usecounter{figcap}
\labelwidth1.6cm \leftmargin2.5cm \labelsep0.4cm \itemsep0ex plus0.2ex }

\item Polar angle dependence of the alignment polarization of $b$-quarks
on the $Z$-peak. Shown are Born term results (dotted) and full
$O(\alpha_s)$ results (solid).

\item Helicity coordinate system defining the polar angle $\theta$ and
the azimuthal angle $\phi$. $\Lambda_b$ is at origin. Decay plane is in the
$(z,x)$-plane. $z$-axis is defined by momentum $\vec p_e$ of the electron
(system I; $(\vec p_X)_x\ge 0$) or momentum $\vec p_X$ of the recoiling
hadron system (system II; $(\vec p_e)_x\ge 0$).

\item Transversity coordinate system defining the polar angle $\theta_T$
and the azimuthal angle $\phi_T$. Decay plane is in the $(x,y)$-plane.
$x$-axis is defined by $\vec p_e$ (system I) and $\vec p_X$ (system II).

\end{list}


\begin{thebibliography}{99}
\bibitem{strass1}The ALEPH Collaboration, CERN preprint PPE/95-156
\bibitem{strass2}J.G.~K\"orner, A.~Pilaftsis and M.M.~Tung,
  Z.~Phys. {\bf C63} (1994) 575;\\
  M.M.~Tung, Phys.~Rev.\ {\bf D52} (1995) 1353;\\
  S.~Groote, J.G.~K\"orner and M.M.~Tung,\\
  Mainz preprint MZ-TH/95-09, hep-ph/9507222, to be published in Z.~Phys.~C
\bibitem{strass3}S.~Groote and  J.G.~K\"orner,
  Mainz preprint MZ-TH/95-17, hep-ph/9508399
\bibitem{strass4}S.~Groote, J.G.~K\"orner and M.M.~Tung,
  Mainz preprint MZ-TH/95-19
\bibitem{strass5}F.E.~Close, J.G.~K\"orner, R.J.N.~Phillips and
  D.J.~Summers,\\ J.~Phys.\ {\bf G18} (1992) 1716
\bibitem{strass6}J.~Chay, H.M.~Georgi and B.~Grinstein,
  Phys.~Lett.\ {\bf B247} (1990) 399;
  I.I.~Bigi, M.~Shifman, N.G.~Uraltsev and A.I.~Vainshtein,
  Phys.~Rev.~Lett.\ {\bf 71} (1993) 496
\bibitem{strass7}A.~Manohar and M.B.~Wise,
  Phys.~Rev.\ {\bf D49} (1994) 1310
\bibitem{strass8}A.F.~Falk and M.E.~Peskin,
  Phys.~Rev.\ {\bf D49} (1994) 3320
\bibitem{strass9}A.~Czarnecki, M.~Jezabek, J.G.~K\"orner and J.H.~K\"uhn,\\
  Phys.~Rev.~Lett.\ {\bf 73} (1994) 384;\\
  A.~Czarnecki and M.~Jezabek, Nucl.~Phys.\ {\bf B427} (1994) 3
\bibitem{strass10}C.~Diaconu, J.G.~K\"orner, D.~Pirjol and M.~Talby,
  Mainz preprint MZ-TH/95-28
\bibitem{strass11}C.~Diaconu, J.G.~K\"orner, D.~Pirjol and M.~Talby,
  to be published
\bibitem{strass12}S.~Balk, J.G.~K\"orner and D.~Pirjol,
  Mainz preprint MZ-TH/95-21
\bibitem{strass13}M.~Gremm, G.~K\"opp and L.M.~Sehgal,
  Phys.~Rev.\ {\bf D52} (1995) 1588
\bibitem{strass14}G.~Bonvicini and L.~Randall,
  Phys.~Rev.~Lett.\ {\bf 73} (1994) 392
\end{thebibliography}
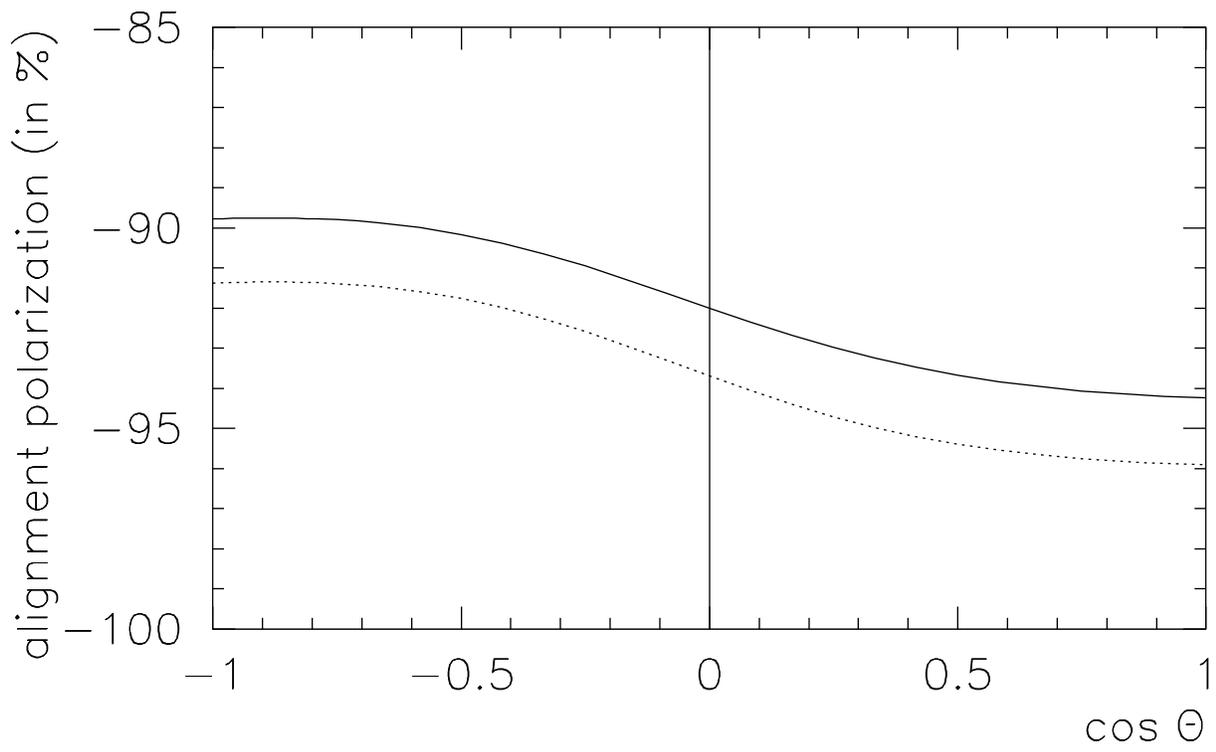
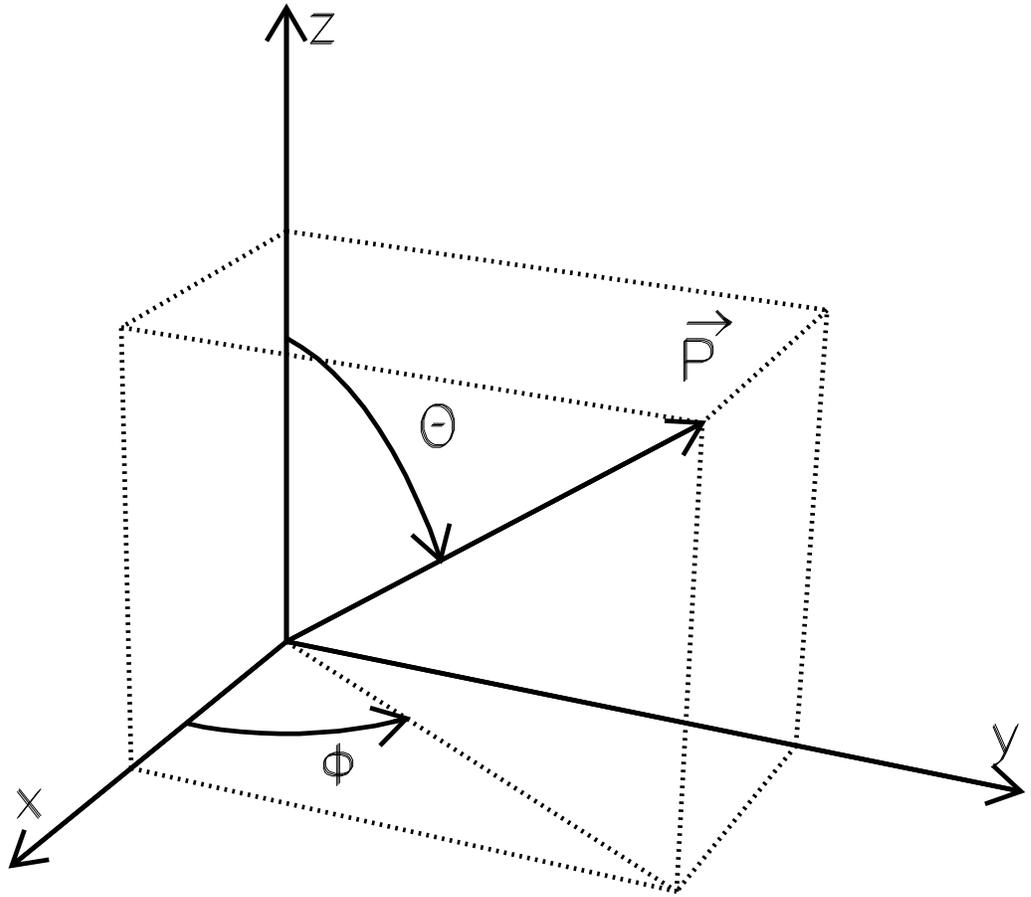
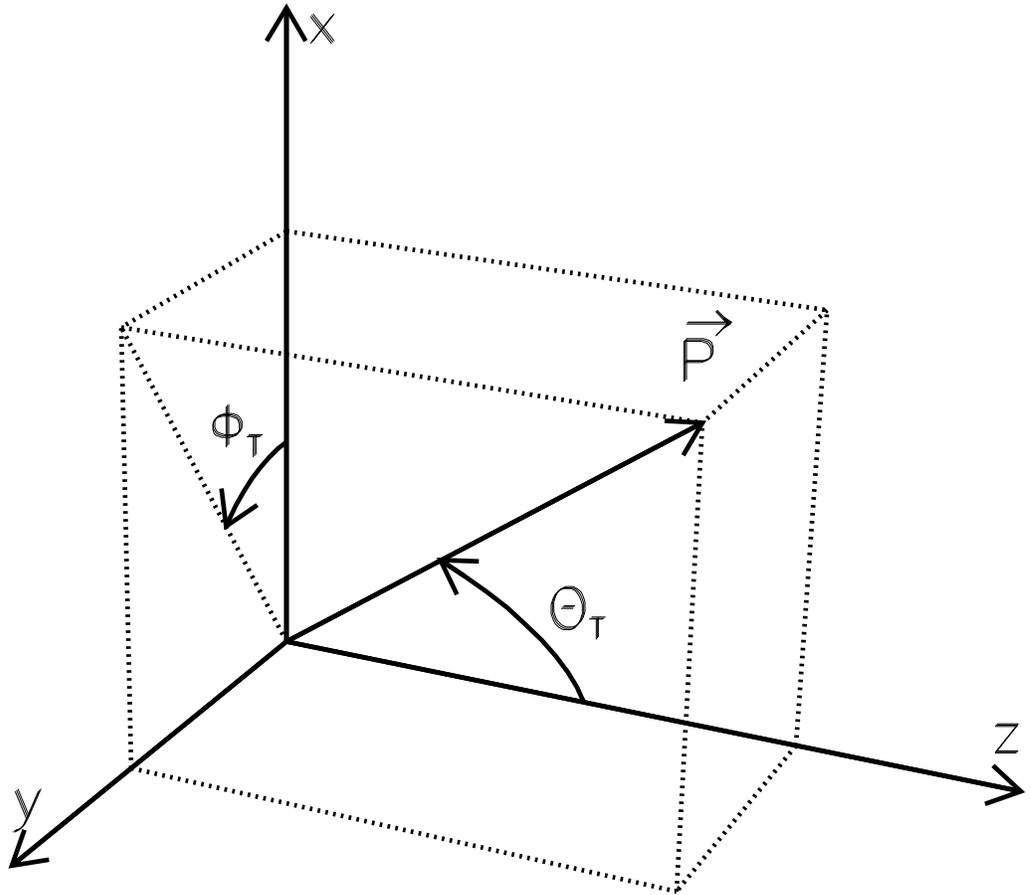
\end{document}